\documentclass[authoryear]{elsarticle}
\usepackage{times} 
\usepackage{natbib}
\usepackage[utf8]{inputenc}
\usepackage{amsmath}
\usepackage{amsfonts}
\usepackage{hyperref}
\usepackage{graphicx}
\usepackage{setspace}
\makeatletter
\newcommand\footnoteref[1]{\protected@xdef\@thefnmark{\ref{#1}}\@footnotemark}
\usepackage{ragged2e}
\justifying
\title{DFT study of five-membered ring PAHs}
\author
{
Gauri Devi$^{1}$,
Mridusmita Buragohain$^{1}$\mbox{*},
Amit Pathak$^{2}$, 
\vspace*{10pt}\\
$^{1}$Department of Physics, Tezpur University, Tezpur 784\,028, India, 
ms.mridusmita@gmail.com, \mbox{*}Corresponding author\\
$^{2}$Department of Physics, Banaras Hindu University, Varanasi 221 005, India 
(amitpah@gmail.com)}

\begin{document}
\begin{frontmatter}
\begin{abstract}
This work reports a ‘Density Functional Theory’ (DFT) calculation of  
PAH molecules with a five-member ring to determine the expected region of 
infrared 
features. It is highly possible that fullerene molecule might be originated 
from five-membered ring PAH molecules in the ISM. Effect of ionization and
protonation on five-membered ring PAH molecule is also discussed. 
A detail vibrational analysis of five-membered ring PAH molecule 
has been reported to further compare with observations and to identify any 
observational counterpart.
\end{abstract}

\end{frontmatter}

\section*{INTRODUCTION}
Polycyclic Aromatic Hydrocarbons (PAHs) are suspected to be one of the stable
and largest aromatic compounds present in the interstellar medium (ISM) of the
Milky Way and external galaxies. The ubiquitous presence of PAHs has been 
established
through numerous observations of the unidentified infrared (UIR) emission bands
in the mid-IR \citep[and references therein]{Tielens08}. PAH molecules, being 
stable, can survive acute conditions
of the ISM in its regular as well as in ionized or de-hydrogenated phase. These
properties help to understand the total carbon budget of the universe. 
PAHs, because of their ubiquitous presence, may provide some clues to some of 
the 
hitherto unsolved problems  of astronomical spectroscopy, e.g. the 3.4~$\mu \rm 
m$ 
absorption feature \citep[]{Sellgren94}, the far-UV steep of the interstellar 
extinction curve \citep[]{Li01}, the
217.5 nm  bump \citep[]{Henning98} and the diffuse interstellar
bands \citep[]{Salama99} that appear as absorption
features in the visible region of the interstellar extinction curve. PAHs are 
also believed
to contribute to the understanding of the formation of molecular hydrogen and
other carbon based molecular systems \citep[]{Pauzat11}. PAHs can also 
contribute to the energetics and the
interstellar chemistry. 

A PAH structure basically consists of fused benzenoid rings with de-localized
electrons. In some cases, these molecules may consist of five-membered ring with 
a
planar (fluoranthene, C$_{16}$H$_{10}$) or a non-planar structure
(corannulene, C$_{20}$H$_{10}$). PAH molecules with five-member ring have
significant interstellar implications in view of the fact that non-planar 
systems belong to polar molecules and are involved with rotational transitions
\citep[]{Pilleri09}. Recently discovered C$_{60}$ and C$^+_{60}$ 
\citep[]{Sellgren10, Campbell15, Ehrenfreund15}
in the reflection nebula NGC 7023 have attracted attentions towards 
five-membered ring PAHs because of
their possibility to help formation of fullerenes or vice versa. In the top-down 
model of 
formation of fullerenes from a PAH molecule, intermediate is a five-membered 
ring PAH \citep[]{Berne12, Mackie15} 
which might show important spectral characteristics
in a UV irradiated source. C$^+_{60}$ are originally proposed to
be the carrier of Diffuse Interstellar Bands (DIBs) in the near-infrared 
\citep[]{Campbell15, Ehrenfreund15}, 
but still awaits its confirmation because of the non-availability of a suitable 
gas phase
spectrum. High resolution spatially resolved spectra of these sources might also
show the spectral signatures of PAHs with five-member rings. Previous studies on 
interstellar PAHs 
reveal that large PAH cations undergo fragmentation through rapid H-loss, 
leaving the molecule
in a fully dehydrogenated form \citep[]{Ekern98, Zhen14}. The process of 
fragmentation depends on various factors including 
the size, compactness and the binding energy of the molecules. Non-compact, 
linear and comparatively small size PAH 
molecules are likely to lose a C$_{2}$ unit forming a pentagon ring, even before 
all the peripheral H atoms are lost 
\citep[]{Ekern98, West14}.

Motivated by the facts that the spectra of C$_{60}$ and C$^+_{60}$ have close
connections with the PAHs with five-member rings, this study proposes to use
Density Functional Theory (DFT) to understand the spectroscopic properties of 
the different forms
of PAHs with five-member rings along with their protonated and cationic
counterparts.    

\section*{COMPUTATIONAL APPROACH}
Density Functional Theory (DFT) is commonly used to study large molecules as
it is computationally efficient, viable and the results closely match those
of experiments. Use of this computational methods helps to calculate the 
harmonic frequencies
and intensities of vibrational modes of PAHs in various forms including size,
composition and charge states \citep[]{Langhoff96, Bauschlicher97, Bausch97b, 
Langhoff98,
Hudgins01, Hudgins04a, Pathak05, Pathak06, Pathak07, Pathak08, Pauzat11-a, 
Candian14, Mridu18}. The
present work uses DFT in combination with a B3LYP functional, along with the
6-311G** basis set used for the purpose of optimization of the molecular 
structures  of
PAHs. The optimized geometry has been used to obtain the vibrational
frequencies of various modes of PAHs. The use of the large basis set such as the 
6-311G**
has certain advantages in reducing the overestimation compared to the smaller
basis sets and shows good agreement with experiment. But it is also beset with 
one
disadvantage, \textit{viz.}, it does not support the use of the single
scaling-factor for the entire range of the vibrational modes 
\citep[]{Langhoff96}. Upon
comparing the theoretical frequencies with laboratory spectroscopic
experimental data, three different scaling factors have been determined.The
scaling factors obtained are 0.974 for the C-H out-of-plane (oop) mode, 0.972 
for the C-H in-plane and C-C
stretching modes and 0.965 for the C-H stretching mode \citep[]{Mridu15, 
Mridu16}. Necessary calibration
for the selected PAH variants are incorporated prior to obtaining the spectra. 
The PAH samples selected for the present works include Corannulene
(C$_{20}$H$_{10}$), Fluoranthene (C$_{16}$H$_{10}$), Benzo-ghi-Fluoranthene 
(C$_{18}$H$_{10}$), Benzo[a]aceanthrylene (C$_{20}$H$_{12}$) with their pentagon 
ring 
at the centre, Azulene (C$_{10}$H$_8$), Acenaphthylene (C$_{12}$H$_8$), Fluorene 
(C$_{13}$H$_{10}$) and Benzo-ghi-Fluoranthene with pentagon ring at the edge. 
The normalized or the relative intensities 
(Int$\rm_{rel}$\footnote{Int$\rm_{rel}$
=$\frac{\rm{absolute~intensity}}{\rm{maximum~absolute~intensity}}$}) and  the 
computed wavelengths of bands are 
used to plot a Gaussian profile with the FWHM of 30 cm$^{-1}$. We have used 
GAMESS \citep[]{Schmidt93} 
and QChem for our calculation \citep[]{YSaho15}.

\section*{RESULTS AND DISCUSSION}
\begin{figure*}
\includegraphics[width=12cm,height=10cm]{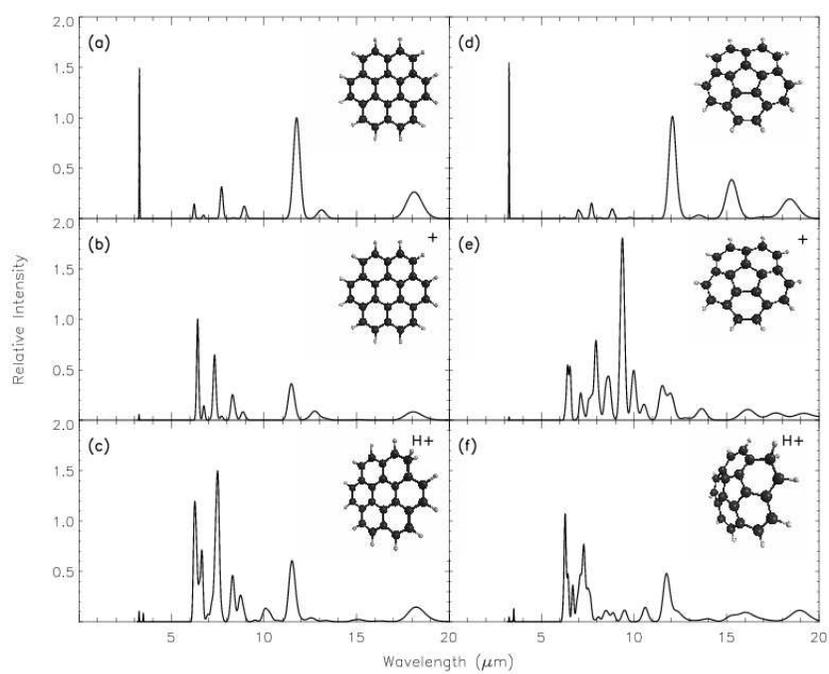}
\caption{Theoretical IR spectra of (a) coronene (C$_{24}$H$_{12}$), (b) coronene 
cation (C$_{24}$H$_{12}^+$), (c) protonated coronene (HC$_{24}$H$_{12}^+$), (d) 
corannulene (C$_{20}$H$_{10}$), (e) corannulene cation (C$_{20}$H$_{10}^+$), 
(f) protonated corannulene (HC$_{20}$H$_{10}^+$)}
\label{fig1}
\end{figure*}

In Fig.~\ref{fig1}, the IR spectra of neutral coronene (C$_{24}$H$_{12}$) 
and neutral corannulene (C$_{20}$H$_{10}$) are compared with their cationic and 
protonated forms. 
The obtained features are the characteristics of various vibrational modes in 
the respective molecules. Coronene is a compact and very stable PAH molecule due 
to
delocalization of electrons and the aromatic bondings between all adjacent 
atoms. 
Corannulene is a five-membered ring PAH with a comparatively less symmetric 
structure, non planer and a close member to coronene. In neutral coronene, the 
dominant features appear at $\sim$3.3~$\mu \rm m$ and 
$\sim$12~$\mu \rm m$ due to C$-$H stretching and C$-$H out of plane vibration 
respectively. 
Ionization of a PAH molecule results in a decrease in the intensity of the 
3.3~$\mu \rm m$ with an increase of 
features in the 6-10~$\mu \rm m$ region \citep[]{Tielens08}. This is indeed
seen for a coronene cation as shown in Fig.~\ref{fig1}b. Protonated coronene 
shows a similar pattern as that of a
coronene cation, but with an increase in the total intensity of features  due to 
lower symmetry of the structure. 
In addition, a 3.4~$\mu \rm m$ feature due to the stretching of 
the aliphatic C$-$H bond is also present for a protonated coronene that is 
observationally detected in various sources in the ISM.  

Neutral corannulene (Fig.~\ref{fig1}d) shows all the features that are also 
obtained for a coronene molecule with an additional overlapping out of plane 
feature at 15.3~$\mu \rm m$ (integrated Int$\rm_{rel}$~0.39). Upon ionization, 
corannulene cation 
not only shows an increase in the intensity of 6-10~$\mu \rm m$ region, but a 
tremendous
increase in the number of features (Fig.~\ref{fig1}e). This characteristic is 
not observed from the cation of a 
regular six-membered ring PAH molecule, for example: coronene. These intense 
features arise due to a combination of C$-$C stretching and C$-$H in plane 
bending in which pentagon ring also takes part. However, a direct effect of 
pentagon ring is rarely present. In this 6-10~$\mu \rm m$ region features of a 
corannulene cation, the most intense
feature appears at 9.4~$\mu \rm m$ (Int$\rm_{rel}$~1). 
Beyond 10~$\mu \rm m$, the features are mostly characteristics of C$-$H out of 
plane vibrational modes. 
Protonation usually results in breaking of the symmetry of the molecule which 
eventually leads to
more number of features with significant intensity. However, protonated 
corannulene shows 
comparatively less intense features (Fig.~\ref{fig1}f) compared to its ionized 
and neutral forms in spite of being protonated. The 3.4~$\mu \rm m$ feature is 
indeed present in the IR spectra of protonated corannulene.
The complete list of wavelength and Int$\rm_{rel}$ of the modes are given as 
supplementary material to the paper.

\begin{figure*}
\centering
\includegraphics[scale=.62]{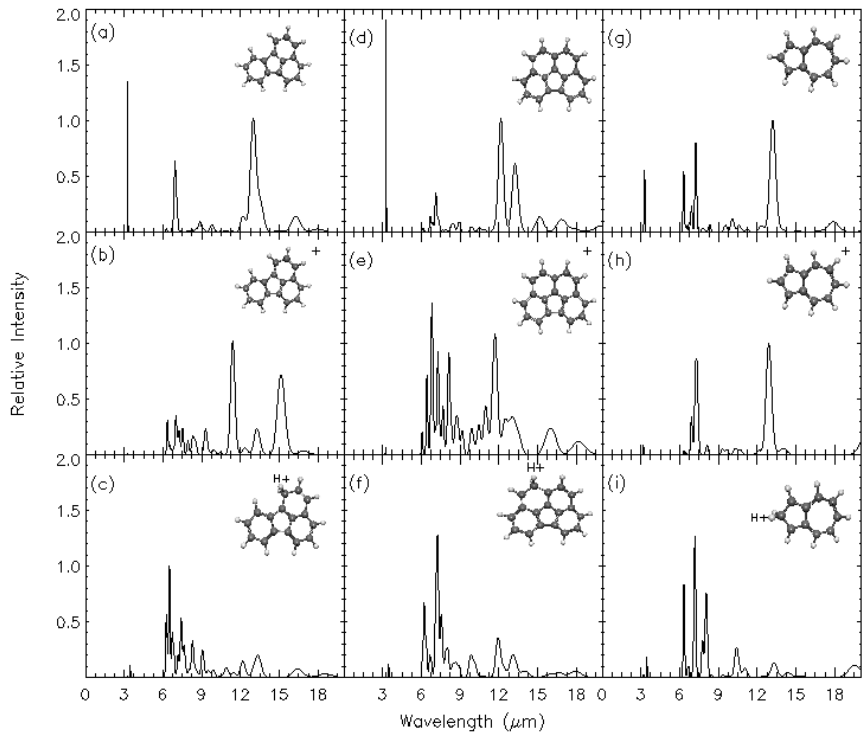}
\caption{Theoretical IR spectra of \hspace{1mm} (\textit{a}) Fluoranthene 
(C$_{16}$H$_{10}$),\hspace{1mm} (\textit{b}) Fluoranthene cation 
(C$_{16}$H$_{10}^+$), \hspace{1mm} 
(\textit{c}) protonated Fluoranthene (HC$_{16}$H$_{10}^+$), \hspace{1mm} 
(\textit{d}) Benzo-ghi-Fluoranthene (C$_{18}$H$_{10}$), \hspace{1mm} 
(\textit{e}) Benzo-ghi-Fluoranthene cation (C$_{18}$H$_{10}^+$), \hspace{1mm} 
(\textit{f}) protonated Benzo-ghi-Fluoranthene (HC$_{18}$H$_{10}^+$), 
\hspace{1mm}
(\textit{g}) Azulene (C$_{10}$H$_8$), \hspace{1mm} (\textit{h}) Azulene cation 
(C$_{10}$H$_8^+$) \hspace{1mm}and \hspace{1mm}(\textit{i}) protonated Azulene 
(HC$_{10}$H$_8^+$).}
\label{fig2}
\end{figure*}

We are also considering comparatively smaller PAHs for our study as smaller PAHs 
are more prone to form a 
pentagon ring followed by a loss of C$_2$ unit as compared to large size PAHs. 
The obtained IR spectra are shown
in Fig.~\ref{fig2}. 
Like any other neutral PAHs, neutral Fluoranthene, neutral 
Benzo-ghi-Fluoranthene and neutral Azulene 
show intense features for C$-$H stretching (3~$\mu \rm m$ region) and C$-$H out 
of plane 
(beyond 10~$\mu \rm m$ region). 
6-10~$\rm \mu m$ region is comparatively less intense for neutral 
Benzo-ghi-Fluoranthene as expected 
whereas Fluoranthene and Azulene in their neutral forms show moderately intense 
6-10~$\rm \mu m$ region. 
Ionization usually enhances features in the the 6-10~$\mu \rm m$ region.
In our study however, only cationic Benzo-ghi-Fluoranthene seems to show this 
behavior with an exception
for Fluoranthene and Azulene.
Ionization of Fluoranthene though leads to an increasing number of features, the 
intensity of the 
individual features seem to be less as compared to its neutral counterpart 
(Fig.~\ref{fig2}b). 
Also, the most intense feature is present at 11.4~$\mu \rm m$ 
(Int$\rm_{rel}\sim$~1) 
due to C$-$C stretching of the central five member ring in combination with 
C$-$C$-$C in plane modes. Another C$-$C$-$C in plane feature appears at 
15.1~$\mu \rm m$ (Int$\rm_{rel}\sim$~0.68).
variation in intensity from that of neutral Azulene.
region). However, cationic Fluoranthene shows
Ionization of Fluoranthene though leads to an increasing number of features, the 
intensity of the 
(Fig.~\ref{fig2}b). 
Protonated form of Fluoranthene shows the expected characteristics; 
\textit{i.e}., less intense 
3~$\mu \rm m$ region with significant 6-10~$\mu \rm m$ region. Cationic 
Benzo-ghi-Fluoranthene 
shows a rich 6-10~$\mu \rm m$ spectra with the most intense feature appearing 
at 6.8~$\mu \rm m$ (Int$\rm_{rel}\sim$~1). These are inherent of C$-$C 
stretching and C$-$H in plane vibrational modes, with an involvement from both 
five and six member rings. In addition, an overlapping 10-13~$\mu \rm m$ region 
is present due to C$-$H out of plane with a partial contribution from C$-$C$-$C 
in plane vibrational modes. Protonated 
Benzo-ghi-Fluoranthene shows a similar spectral effect as that of protonated 
Corannulene, 
where the number of spectral features in the 6-10~$\mu \rm m$ region decreases 
with 
a decrease in the total intensity in comparison to its respective cations. 
Another striking difference for protonated Benzo-ghi-Fluoranthene is that the 
C$-$H out of plane 
intensity at 11-13~$\mu \rm m$ decreases, whereas the same feature appears 
significantly for its neutral and cationic form. 
In the 6-10~$\mu \rm m$, cationic Azulene shows a slight decrease in numbers 
with insignificant variation in intensity from that of neutral Azulene. 
Protonated Azulene shows rich spectra in 6-10~$\mu \rm m$ region; the 
most intense feature appears at 7.2~$\mu \rm m$ (Int$\rm_{rel}\sim$~1) due to 
C$-$C stretching of the pentagon ring in combination with C$-$C and C$-$H in 
plane vibrational mode. 

\begin{figure*}
\centering
\includegraphics[scale=.62]{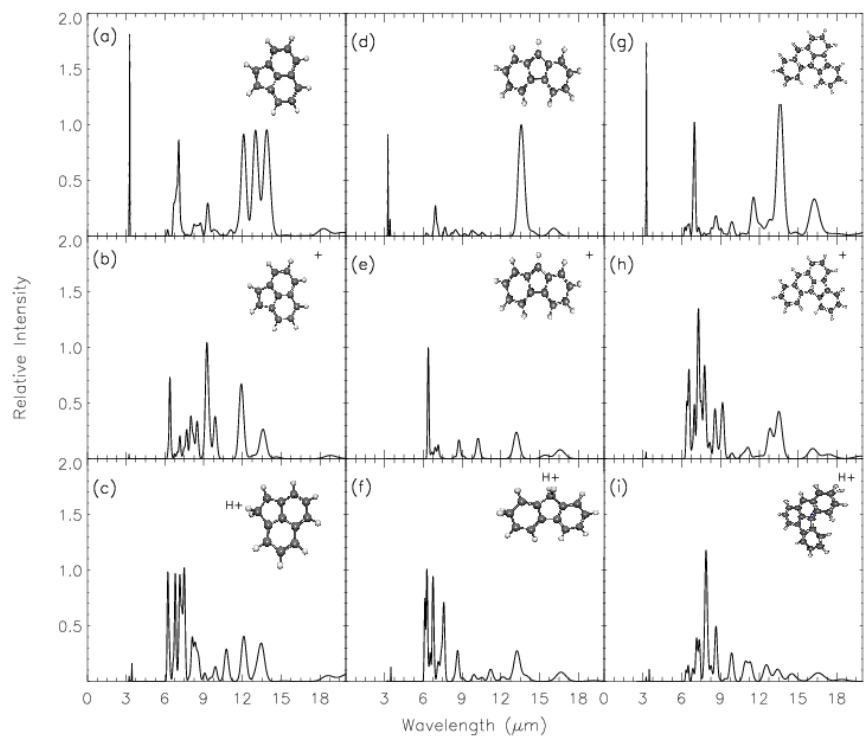}
\caption{Theoretical IR spectra of \hspace{1mm} (\textit{a}) Acenaphthylene 
(C$_{12}$H$_{8}$),\hspace{1mm} (\textit{b}) Acenaphthylene cation 
(C$_{12}$H$_{8}^+$), 
\hspace{1mm} (\textit{c}) protonated Acenaphthylene (HC$_{12}$H$_{8}^+$), 
\hspace{1mm} 
(\textit{d}) Fluorene (C$_{13}$H$_{10}$), \hspace{1mm} (\textit{e}) Fluorene 
cation (C$_{13}$H$_{10}^+$), \hspace{1mm} 
(\textit{f}) protonated Fluorene (HC$_{13}$H$_{10}^+$), \hspace{1mm}
(\textit{g}) Benzo[a]aceanthrylene (C$_{20}$H$_{12}$), \hspace{1mm} (\textit{h}) 
Benzo[a]aceanthrylene cation (C$_{20}$H$_`{12}^+$) \hspace{1mm}and 
\hspace{1mm}(\textit{i}) protonated Benzo[a]aceanthrylene 
(HC$_{20}$H$_{12}^+$).}
\label{fig3}
\end{figure*}
Fig. 3 shows the IR spectra of neutral Acenaphthylene, neutral Fluorene and 
neutral Benzo[a]aceanthrylene. The obtained spectra show similar features as 
that of 
Figs. 2a, 2d and 2g. However neutral Acenaphthylene and neutral 
Benzo[a]aceanthrylene show additional
overlapping out of plane features at 12.1~$\mu \rm m$ and 
11.4~$\mu \rm m$ (Int$\rm_{rel}\sim$~1) respectively. Although ionization has 
increased the number of features in the 6-10~$\mu \rm m$ region,  the increase 
in 
the intensity of the features in 6-10~$\mu \rm m$ is not appreciable as compared 
to its neutral counterpart in Fig. 3a; most intense features appear at 9.2~$\mu 
\rm m$ 
and 6.4~$\mu \rm m$ (Int$\rm_{rel}\sim$~1) due to the C$-$C stretching and C$-$H 
in plane mode of vibration. Protonated Acenaphthylene shows prominent features 
in 6-10~$\mu \rm m$ 
region with three most intense features at 6.2~$\mu \rm m$, 6.8~$\mu \rm m$ and 
7.1~$\mu \rm m$ (Int$\rm_{rel}\sim$~1) due to C$-$C stretching mode of pentagon 
ring in 
combination with C$-$H in plane mode of vibration. Cationic Fluorene shows no 
significant 6-10~$\mu \rm m$ in spite of being ionised. Most intense feature in 
this case appear 
at 6.4~$\mu \rm m$ (Int$\rm_{rel}\sim$~1) due to the C$-$C stretching and C$-$H 
in plane mode of vibration. Protonation of Fluorene has increased both the 
number and intensity 
of the features with most intense features at 6.3~$\mu \rm m$, 6.1~$\mu \rm m$ 
and 6.7~$\mu \rm m$ (Int$\rm_{rel}\sim$~1) due to C$-$C stretching of the 
pentagon ring, 
6-member rings and C$-$H in plane vibration mode. Cationic Benzo[a]aceanthrylene 
shows a tremendous increase in the number and intensity of the features in 
6-10$\mu \rm m$ 
region; the most intense feature appears at 7.1~$\mu \rm m$ 
(Int$\rm_{rel}\sim$~1) due to C$-$C stretching of the pentagon ring in 
combination with C$-$H in plane 
vibration mode. Protonated Benzo[a]aceanthrylene has shown comparatively less 
intensity features even though the number of features are increased. The most 
intense 
feature appears at 7.7~$\mu \rm m$ due to the C$-$C stretching of pentagon ring 
combined with C$-$C$-$C in plane vibration mode.

\begin{figure*}
\centering
\includegraphics[width=12cm,height=10cm]{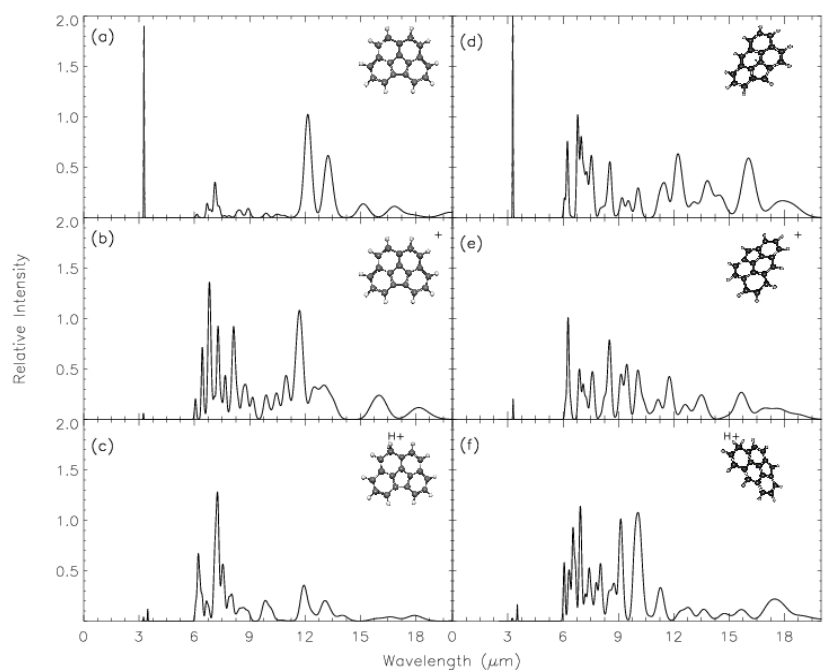}
\caption{Theoretical IR spectra of Benzo-ghi-Fluoranthene (C$_{18}$H$_{10}$) 
with the pentagon ring at the centre: (a) neutral form (C$_{18}$H$_{10}$), 
\hspace{1mm} (b) cationic form (C$_{18}$H$_{10}^+$) and  \hspace{1mm} (c) 
protonated form (HC$_{18}$H$_{10}^+$). The theoretical IR spectra for the 
neutral, cationic 
and protonated forms of the Benzo-ghi-Fluoranthene (C$_{18}$H$_{10}$) when the 
pentagon ring is at the edge are given in (d), (e) and (f) respectively.}
\label{fig4}
\end{figure*}

As observed from the IR spectra obtained in Fig. 4, it is clear that neutral 
Benzo-ghi-Fluoranthene with the pentagon ring at the edge shows a rich 6-10~$\mu 
\rm m$ 
region as compared to the neutral Benzo-ghi-Fluoranthene with the pentagon ring 
at the centre. These features are due to C$-$C stretching and C$-$H in plane 
vibrational modes
with contribution from both five and six member rings; the most intense feature 
appears at 6.7~$\mu \rm m$ (Int$\rm_{rel}\sim$~1). With ionization, 
Benzo-ghi-Fluoranthene
with the pentagon ring at the centre shows a rich 6-10~$\mu \rm m$ region 
similar to the cationic Benzo-ghi-Fluoranthene with pentagon ring at the edge. 
The most 
dominant feature of cationic Benzo-ghi-Fluoranthene with pentagon ring at the 
centre appears at 6.9~$\mu \rm m$ (Int$\rm_{rel}\sim$~1) due to C$-$C in plane 
stretching 
in combination with C$-$H in plane mode of vibration. In ionized 
Benzo-ghi-Fluoranthene with the pentagon ring at the edge, there is a slight 
increase in the intensity of the 
features, but the number of features are almost the same. However, reverse is 
the situation in the case of protonation. Protonated Benzo-ghi-Fluoranthene with 
the pentagon
ring at the edge shows a tremendous number of features and increase in intensity 
of the features in 6-10~$\mu \rm m$  region. The most intense features are at 
6.4~$\mu \rm m$, 
 6.8~$\mu \rm m$, 8.9~$\mu \rm m$ and 9.6~$\mu \rm m$ (Int$\rm_{rel}\sim$~1) due 
to C$-$C in plane stretching of the pentagon ring and six member ring.

We are providing the complete list of position and intensity of bands as 
supplementary material to the paper.

\section*{CONCLUSION}
Five-membered ring PAH molecules are important, particularly in ionized forms in 
the ISM in view of its possible relation 
in the formation of C$_{60}^+$. We have carried out DFT study for some selected 
five-membered ring PAH molecules to
understand its spectral characteristics and to find correlation with the 
observed AIBs, if any. Five-membered ring PAHs
do not show any specific extraordinary feature which we can assign to be 
exclusively coming from the pentagon ring.
However, indirect effect of pentagon ring is indeed present which are evident 
particularly for cations of five-membered
ring PAHs. Any PAH cations are expected to produce intense features in the 
6-10~$\mu \rm m$ region. 
These features grow even more intense for a five-membered ring PAH cations. This 
might be an indirect effect 
coming from the involvement of a pentagon ring. Our study shows some specific 
spectral characteristics for 
a neutral five-membered ring PAH (Benzo-ghi-Fluoranthene with the pentagon ring 
at the edge), that can probably be attributed to the location of 
the pentagon ring present in it. Interestingly, protonated forms of 
five-membered ring PAHs show moderate 6-10~$\mu \rm m$ region whereas for a PAH 
containing only
six-member ring, a broad and intense 6-10~$\mu \rm m$ region is seen. In such a 
scenario,  6-10~$\mu \rm m$ region
stands as a strong tool that might be used for the observational search of 
five-membered ring PAHs in the ISM. 
A detail study in terms of size and shape is indeed required to draw a concrete 
conclusion. 
This study might also be helpful to understand the formation of interstellar 
fullerenes from five-membered ring PAHs.  

\section*{Acknowledgements}
AP acknowledges financial support from ISRO Respond grant (ISRO/RES/2/401/15-16) 
and DST EMR grant, 2017. 
AP thanks the Inter-University Centre for Astronomy and Astrophysics, Pune for 
associateship. AP and MB acknowledge
financial support from DST JSPS grant (DST/INT/JSPS/P-238/2017).
\section*{REFERENCES}
\def\aj{AJ}%
\def\actaa{Acta Astron.}%
\def\araa{ARA\&A}%
\def\apj{ApJ}%
\def\apjl{ApJ}%
\def\apjs{ApJS}%
\def\ao{Appl.~Opt.}%
\def\apss{Ap\&SS}%
\def\aap{A\&A}%
\def\aapr{A\&A~Rev.}%
\def\aaps{A\&AS}%
\def\azh{AZh}%
\def\baas{BAAS}%
\def\bac{Bull. astr. Inst. Czechosl.}%
\def\caa{Chinese Astron. Astrophys.}%
\def\cjaa{Chinese J. Astron. Astrophys.}%
\def\icarus{Icarus}%
\def\jcap{J. Cosmology Astropart. Phys.}%
\def\jrasc{JRASC}%
\def\mnras{MNRAS}%
\def\memras{MmRAS}%
\def\na{New A}%
\def\nar{New A Rev.}%
\def\pasa{PASA}%
\def\pra{Phys.~Rev.~A}%
\def\prb{Phys.~Rev.~B}%
\def\prc{Phys.~Rev.~C}%
\def\prd{Phys.~Rev.~D}%
\def\pre{Phys.~Rev.~E}%
\def\prl{Phys.~Rev.~Lett.}%
\def\pasp{PASP}%
\def\pasj{PASJ}%
\def\qjras{QJRAS}%
\def\rmxaa{Rev. Mexicana Astron. Astrofis.}%
\def\skytel{S\&T}%
\def\solphys{Sol.~Phys.}%
\def\sovast{Soviet~Ast.}%
\def\ssr{Space~Sci.~Rev.}%
\def\zap{ZAp}%
\def\nat{Nature}%
\def\iaucirc{IAU~Circ.}%
\def\aplett{Astrophys.~Lett.}%
\def\apspr{Astrophys.~Space~Phys.~Res.}%
\def\bain{Bull.~Astron.~Inst.~Netherlands}%
\def\fcp{Fund.~Cosmic~Phys.}%
\def\gca{Geochim.~Cosmochim.~Acta}%
\def\grl{Geophys.~Res.~Lett.}%
\def\jcp{J.~Chem.~Phys.}%
\def\jgr{J.~Geophys.~Res.}%
\def\jqsrt{J.~Quant.~Spec.~Radiat.~Transf.}%
\def\memsai{Mem.~Soc.~Astron.~Italiana}%
\def\nphysa{Nucl.~Phys.~A}%
\def\physrep{Phys.~Rep.}%
\def\physscr{Phys.~Scr}%
\def\planss{Planet.~Space~Sci.}%
\def\procspie{Proc.~SPIE}%
\let\astap=\aap
\let\apjlett=\apjl
\let\apjsupp=\apjs
\let\applopt=\ao
\bibliographystyle{elsarticle-harv} 
\bibliography{mridu}

\end{document}